\documentstyle[12pt]{article}
\makeatletter
\begin{document}
\setlength{\baselineskip}{3ex}
\pagestyle{plain}
\pagenumbering{arabic}
\setcounter{page}{1}

\begin{center}
{\bf \Large On the Vacuum Theta Angle in Yang-Mills Field Theories}
\\
\vspace{1 cm}
{\bf \sf   Liang-Xin Li and Bing-Lin Young}\\
\vspace{1 cm}
Department of Physics and Astronomy, Iowa State
University\\
Ames, Iowa 50011\\
May 12,  1994\\
\end{center}
\vspace{3cm}
\begin{abstract}
Recently, MIT group found a numerical solution to the non-abelian gauge field
theory. This solution is shown to contain non-integer topological numbers. By
using this fact, V.V.Khoze has proved that the vacuum theta-angle is zero in
non-abelian gauge field theories. Here, we study the vacuum structure by a
complimentary way-its physical respect. The key point is that MIT solution has
an infinite action, which means there is no tunnelling between the non-integer
topological field configrations although the continuous fractional gauge
transformations which is allowed in Minkowski space
 eliminate the topological distinction between different
winding number sections. Using this fact,we prove that the $\theta$ is
zero at a semiclassical level.
\end{abstract}

\newpage
In the classical picture, the vacuum of the non-abelian gauge theories is
a many-state $\theta$ vacuum[1]. Only the integer winding number gauge
transformations are allowed[2]. Correspondingly the vacuum states are the
integer
topological number field configurations which are connected by the instanton
tunnelling[3]. This leads to the periodic picture in the one instanton sector.
This periodicity contains an intrinsic $\theta$ angle which is similiar to the
bloch wave number in the periodic potential case. However, all of this is in
Euclidean space. Recently, MIT group[4] find that in Minkowski space, the
continuous fractional gauge transformations are allowed. This leads to the
collapse of the periodic picture. V.V.Khoze[5] showed that there is a unique
vacuum actually, and the $\theta$ angle is zero.

In this paper, we look at this in a different way. We treat the vacuum as a
periodic potential quantum problem and try to get the relation between the
bloch wave number and the $\theta$ angle. Then we consider the solution [4]
's effect on this picture. We prove that the $\theta$ angle is zero only if
the infinite action of the solution means the tunneling between the fractional
winding number configurations are zero.

    Since BPST[6] found the instanton solutions in Euclidean space, the vacuum
structure is considered as the periodic $\theta$ Vacuum: Only the integer
winding number field configurations are allowed in the vacuum, and the
instanton is the most effective tunnelling path. The tunneling probability is
really small[7] due to the topological suppress. Correspondingly the $\theta$
angle is very small. To look at the origin of the $\theta$, we briefly review
the vacuum picture:
 In the non-abelian gauge field theory:
$$S=-\frac{1}{2g^2}{\displaystyle \int}d^4xtr(F^{\mu\nu}F_{\mu\nu})$$
where the field strength
$F_{\mu\nu}=\partial_{\mu}A_{\nu}-\partial_{\nu}A_{\mu}-i[A_{\mu},A_{\nu}]$,
and $A_{\mu}=A_{\mu}^a\frac{\tau_a}{2}$, where $\tau_a$,$a=1,2,3$ are the
pauli matrix. As far as the topological properties of the vacuum structure are
concerned, the $SU(2)$ theory is enough. g is the coupling constant.

First of all,in Euclidean space, we have many vacuum
states[8]: $|n>$, $n=0$, $\pm 1$,$\pm 2$,....$\pm \infty$, where n is the
winding
number of the gauge transformatioms $U_n$. So the vacuum
$$A_{\mu}^n=U_n(x)\partial_{\mu}U_n^{-1}(x)$$ is classified by the
Pontriyagin index $\nu[A]$
\begin{equation}
\nu[A]=\frac{1}{16\pi^2}{\displaystyle
\int}d^4xtr(\tilde{F}_{\mu\nu}F^{\mu\nu})
\end{equation}
where
$\tilde{F}_{\mu\nu}=\frac{1}{2}\epsilon^{\mu\nu\alpha\beta}F_{\alpha\beta}$ is
the dual fields in
Euclidean space. The equation (1) can be reduced to a surface integral
which equals the change of the  winding number in the boundary. In the vacumm
section $\nu[A]$ is the winding number of the corresponding gauge
transformations.
The winding number $\mu[U]$ is defined as:
\begin{equation}
\mu[U]=-\frac{1}{24\pi^2}{\displaystyle
\int}^{\infty}_{-\infty}d^3x\epsilon^{ijk}
((U^{+}\partial_iU)(U^{+}\partial_jU(U^{+}\partial_kU) )
\end{equation}
and $\mu[U_n]=n$.
Actually the integrand of (2) is the jocobi matrix of the homotopy mapping from
the group
manifold to the Euclidean space boundary $S^3$:
$$S^3_G\longrightarrow S^3_E$$
If we demand any gauge transformations satisfy
$$U(x)\longrightarrow 1~~~~as~~~~~ |x|\longrightarrow\infty$$
where $|x|^2=x_0^2+x_1^2+x_2^2+x_3^2$.
then the Euclidean space can be compactified as $S^3$ which leads to the above
classification.
So only integer winding numbers are allowed in this picture. And it is easy to
show $U_nU_m=U_{m+n}$. Then in the Hilbert space:
$$|n+m>=g^{U_m}|n>$$
where $g^{U_n}$ is the gauge transformation operator.
$$A_i^{U}=g^{U}A_ig^{U^+}=U^+(\partial_i+A_i)U$$
In the quantization of the theory, $g^U$ is constructed by the field operators.
 However, it is required
that the physical vacuum is a gauge invariant object. We need[3]:
\begin{equation}
g^{U}|vac>_{phys}=e^{-i\mu[U]\theta}|vac>_{phys}
\end{equation}
the explicit construction of $|vac>_{phys}$ is
\begin{eqnarray}
|vac>_{phys}=|\theta>_{vac}\nonumber\\
|\theta>_{phys}=\sum e^{in\theta}|n>
\end{eqnarray}
Here  the $\theta$ angle is an arbitrary phase angle which can't be determined
by
any physical principle.

But the physical space time should be the Minkowski space time. In this case,
the MIT solution serve the role as the instanton in Euclidean space. This
solutions have the vacuum which contains
 continuous fractional winding number configurations instead of the
dicrete integer field. From the equation (2), we know the gauge
transformations are also continuously fractional. From the point of the
homotopy, this implies  the different winding number mapping $U_n$ can be
continuously deformed into each other now. So all the mapping $\{U_n\}$ are in
the same homotopy class by the definition of homotopy. Therefore, there is no
topological distinctions between the mappings. From the equation(2), we
know all the vacuum states are in the same class. The vacuum state is unique
from the point of topology. There is no periodic picture in the Minkowski
space, then the $\theta$ does not exist. In other words, it is zero. This is
the
basic argument of the paper[5].

  However, above is more mathematical way than physical picture.. We look at
this at a different and more physical way. From the classical picture in the
above,, we treat the
many vacuum states as a periodic potential systems in the winding number
space. That is $V(q)=V(q+n)$ with q as the coordinate of winding number.
$V(q)$ is the potential of the gauge field, and n
is a integer. We know that for a periodic potential, the wave function takes
the bloch wave forms[9]:
\begin{eqnarray}
|\psi>&=&e^{ikq}U(q)\nonumber\\
U(q)&=&U(q+n)
\end{eqnarray}
 $U(q)$ is a periodic function in the winding number space. We construct it
as
$$U(q)=\sum U(q+n)$$
Through some algebraic work, we have
\begin{equation}
|\psi>=\sum e^{-ikn^{\prime}}U^{\prime}(n^{\prime})
\end{equation}
where $$U^{\prime}(n^{\prime})= e^{ikn^{\prime}}U(n^{\prime})$$
where $n^{\prime}=q+n$.Here we ignore a total phase term.
 From quantum mechanics, we know that when the
potential is very steep with minimum at integers, then the function is peaked
at the minimum points(like $e^{-\alpha q^2}$). We can approximately treat
the $q^{\prime}$ as zero which is the first minimum point.Now  we have
$$|\psi>=\sum e^{-ikn}|n>$$ Compared with the equation(3), we can identify k
as the $\theta$. From here, we can see that the $\theta$ is the bloch wave
number in the winding number space. In the following example, we can find
that the $\theta$ is actually mixing angle due to the tunelling between
different field configuration.

We proceed the bloch method furtherly by considering a periodic delta
potential which is very close the vacuum picture in the Minkowski space in our
approximation. Actually one can paramatrize the potential and mass as
functions of winding number q[9] by using the exact tunneling path solution.
 Here the detail function is not important.
 Now
we have the potential V(q)
$$V(q)=\lambda\delta(q-n)$$ where n is an arbitrary integer.
the schrodinger equation is
$$ \frac{d^2\psi}{dq^2}+\frac{2m}{{\bar{h}}^2}(E-V(q))\psi=0$$
 from bloch theorem,
the solution of this equation will have the following properties:
$$\psi(q+n)=e^{i\theta n}\psi(q)$$
in the period $0<q<1$, the $\psi(q)$ has the form
$$ \psi(q)=Ae^{i\beta q}+Be^{-i\beta q}$$ where
$\beta=\sqrt{\frac{2mE}{{\bar{h}}^2}}$.
According to the bloch theorem, the wave function in the period $1<q<2$ is
$$\psi(q)=e^{i\theta}(Ae^{i\beta(q-1)}+Be^{-i\beta(q-1)})$$
By using the connection conditions, one can find the non-traivial solutions
are obtained only if the determinant of the coefficients of A and B
vanishes, which gives the conditions:
\begin{equation}
\cos\theta=cos\beta+p\frac{\sin\beta}{\beta}
\end{equation}
where $p=\frac{m\lambda}{{\bar{h}}^2}$. From the equation(7), one can get the
energy  band
spectrum of this approximation of the vacuum structure
Here we are only interested in the case when $\lambda\longrightarrow \infty$.
In this case, to keep the equation (7), we need
$$\sin\beta=0$$
to have finite energy solutions. So
\begin{equation}
\cos\theta=\pm 1
\end{equation}
So $\theta=\pi n$ with n as arbitrary non-zero integers. For the $\beta=0$, we
directly substitute into the determinant to see that it keeps the original
determinat of the coefficients zero.
Then $\beta=0$ is also the solution as the $\lambda\longrightarrow \infty$.
We see  the tunneling goes to zero as the barrier between two different
regions goes to infinity,
 then the mixing $e^{i\theta}$ goes to unity which means there is no mixing.

In the Yang-Mills field theory case, we know the action of MIT solution[4] is
infinite. Also, from the physical meaning of this solution, we can see this
solution is the connection path in physical space time between $\nu=0$ and
$\nu=q$:

The solution is to describle the moving spherical energy shell's
effect on a finite region $r<R$: initially there is nothing in the region
$r<R$ which means $$A_{\mu}=\frac{1}{g}U\partial_{\mu}U^+=0$$

{}From(2), this means  $\nu=0$ at the initial situation. The energy shell is
moving toward this region from
$t=-\infty$, then
after a long time, the energy shell is reflected back to infinite at
$t=\infty$, however the final state of the region $r<R$ is changed to a
different configuration with $\nu=q$, and
$$A_{\mu}=\frac{1}{g}U\partial_{\mu}U^+$$

where $$U=exp(iq\Lambda(r)\frac{\hat{x}\dot \tau}{2})$$
q is an arbitrary fractional number, and the $\Lambda(r)$ is satisfying the
boundary conditions
$$\Lambda(r)|_{r\rightarrow 0}\longrightarrow
0,~~~~~\Lambda(r)|_{r\rightarrow\infty}\longrightarrow 2\pi$$
The winding number of this U(r) is
\begin{equation}
\mu[U]=q-\frac{\sin(2\pi q)}{2\pi}
\end{equation}
by calculating (2).

 From this picture we see clearly the
solution is a quantum path which connects the $\nu=0$ with $\nu=q$.

Nontheless,the action is infinite which actually means there is no connection
in the physical space time. Our key argument is as follows: In the nearby of
this solutin, there are many other quantum path which also have infinite or
very large action. And the transtion probability is proportional to $e^{iS}$
where the S is the action of the path. So the large action means the very fast
oscillation of the probability which eventually is washed out by the fast
oscillation. Then actually there is no transition between these field
configurations. Here we use the infinite potential  barrier between the $\nu=0$
and $\nu=q$ to realize this situation in the winding number space.
In the above, we take the delta function with the
integral height going to infinte to simulate the real case. However the
distance between two potential peaks is not the unity but arbtrarily short due
to the continuous q allowed. To consider this situstion, we change the the
potential to $V(q)=\lambda\delta(q-na)$ with a goes to zero. In this potential,
the
equation  is changed into:
\begin{equation}
\cos\theta=\cos\beta a+p\frac{\sin\beta a}{a\beta}
\end{equation}
So when p goes to infinity, we get
$$\beta a =\pi n$$
with $n=0,\pm 1, \pm 2,...$. Then we take the limit $ a\longrightarrow 0$
and keep the finite energy solutions, one immediately get
$$ \beta a=0$$ only the $n=0$ is allowed. So we get
\begin{equation}
\cos\theta=\pm 1
\end{equation}
So only the $\theta=0$ is physically reasonable.
\vspace{1cm}

Here, this conclusion is only depending on the assumption: periodic potential
in the winding number space  and the infinite energy barrier between
fractional winding number field configurations in the Minkowski space. We
should note this conclusion does not depend on the details of the periodic
potentials, for example the periodic block potential, the periodic harminic
potential. However, we assume the MIT solution is the most effective path in
the above. This point is a still open problem[10]. Right now we  don't know
whether there is other finite action path or not. To proceed this respect, we
need to solve the non-linear field eqautions both in the dual equation case[10]
and the second order differential equation situation[5]. On the other hand,
from
above we know the $\theta$ angle is from the interference and tunnelling
between different sector this is responsible for the smallness of the $\theta$
angle.

\end{document}